\def\beq{ \begin{equation}}
\def\eeq{ \end{equation}}
\def\ba{ \begin{eqnarray}}
\def\ea{ \end{eqnarray}}

\documentstyle[aps,preprint,12pt,epsf]{revtex}
\tightenlines

\title{ On the Hawking Turok solution to the Open Universe wave function}

\author{ W. G. Unruh}
\address{ Program in Cosmology and Gravity of CIAR\\Dept 
Physics and Astronomy\\University 
of B.C.\\ Vancouver, Canada V6T 1Z1}
\begin{document}
\maketitle

\begin{abstract}

Hawking and Turok have recently published a solution to the WKB
``wave-function for the universe" which they claim leads in a natural
way to an open universe as the end point of the evolution for a
universe dominated by a scalar field. They furthermore argue that their
solution a preferred solution under the rules of the game. This paper
will, I hope, clarify their solution and the limits of validity of
their argument.

 \end{abstract}

Hawking and Turok \cite{HT} (hereafter HT) recently published a
fascinating paper in which they claimed to have a solution of the
(complex) Einstein equations which satisfy the rules for a WKB solution
for the ``wave-function of the universe" as laid down by Hartle and
Hawking\cite{HH}, but which has as the future 3-geometry a spatially
open hyperbolic homogeneous universe. Thus this ``instanton" solutions
allows them to calculate the probability that the universe will form as
a homogeneous open universe. Using the anthropic arguments, they also
claimed to find a preferred value for the present day $\Omega$, the
ratio of density to critical density, of 0.01, rather than the value of
1 as usually expected of inflationary models.   This of course is ``too
small" in comparison with experimental evidence, just as the usual
prediction of unity may be ``too large"\cite{Omega}. Linde\cite{Linde}
has also criticized their derivation. However, their derivation was
somewhat elliptic and clarification of what they calculated seems
desirable. This paper will I hope provide such clarification and will
also raise some questions about the
interpretation of their solution.

Let us first review the rules\cite{HH}. The wave function for the universe 
$\Psi(~^3{\cal G})$ is a wave function defined on Euclidean non-singular 
three universes $~^3{\cal G}$. It is defined (at least formally) by means 
of a path integral, where the path integral is to be taken over all four 
geometries, including complex four geometries, which are 4 dimensional 
manifolds whose only boundary is the three geometry of interest. 
This path integral prescription is  poorly defined,  and thus its
principle application has been via a semi-classical approximation. 
One chooses as one's allowed 4 geometries 
only those which extremize the Einstein-matter action, and have the
given three geometry as their only boundary. 
 One hopes that there are only a few, or one, of these solutions.  One
then takes as the approximation to the wave-function just $\Sigma_j
e^{iS_j(^3{\cal G})}$ where $ S_j$ is the action for the $j^{th}$
4-geometry solution. The real part of $S_j$ is then the phase of the
wave function, while the imaginary part gives the probability for that
particular component of the wave function. Since there is no known
normalisation procedure for these wave-functions, such a probability
can best be regarded as a relative probability for evaluating various
solutions (and perhaps various three geometries, although the
normalisation problems remain a difficult problem).

One issue which will arise in the following is the class of solutions
to Einstein's equations which are allowed in this approximation scheme.
Einstein's equations allow  singular solutions. Are those
singularities to be removed from the manifold? Do the singularities
represent a "boundary"? What sort of singularities should be allowed to
be a part of the acceptable solutions? There are clearly a large number
of 4-geometries which will interpolate to the given 3-geometry on the
boundary. What are the smoothness conditions which one should place on
such 4-geometries?
It is certainly true that one can find an extremely large number of WKB
"solutions" to the Einstein  equations if one simply excises those
regions where the Einstein equations are not obeyed and calls them
``singularities".  These solutions can, furthermore, have finite
action. Does one demand that the "singularity" must be a curvature
singularity? Would for example a negative mass Schwartzschild solution
be an acceptable singularity within such an interpolating 4-geometry?
Would the existence of such a negative mass Schwartzschild singularity
in the initial three geometry be a valid three geometry for which one
wishes to determine the probability? Are there criteria by which one
could categorise those classes of singularities which are allowed, and
those which are not?  I raise these concerns, not in order to give a
solution (the simplest solutions is clearly to disallow all
singularities), but because they arise in considering the HT solutions.
The HT solutions contains singularities not only in the interpolating 4
geometry, but also on the three geometry for which one is calculating
the wave-function.

These problems with dealing with singularities are especially difficult
as the interpolating 4 geometries need not be real at all. One cannot
therefor make arguments in which say the space-like of time-like nature
of the singularity plays a crucial role, as these concepts are in
general inapplicable to a complex geometry.

Let us now examine the Hawking Turok ``instanton" in light of this
prescription.  They choose as their universe a homogeneous solution of
Einstein's equations.
\beq
ds^2= d\sigma^2 + b(\sigma)^2(d\psi^2+ sin(\psi)^2 d\Omega^2)
\eeq
where $d\Omega^2= d\theta^2+\sin(theta)^2 d\varphi^2$, the metric on a sphere with
coordinates $\theta,~\varphi$. $b(\sigma)$ is to be taken as the scale
factor for a
 Euclidean solution for a homogeneous universe coupled to a homogeneous
scalar field. The scalar field and the metric solutions are chosen so
that
 at $\sigma=0$ the factor $b(\sigma)=\sigma$. This ensures  that the
point $\sigma=0$ is a regular point in the Euclidean spacetime.  In
general, the solution will be such that $ b(\sigma)$ will increase as a
function of $\sigma$ and then decrease again to zero, but this time with  a singularity where
$b(\sigma)$ goes to zero.

In particular, the evolution of the spacetime is determined by the constraint equation
\beq
6b (\dot b)^2 -6b  +b^3\dot\phi +b^3 V(\phi)=0
\eeq
and the equation for $\phi$, the scalar field  
\beq
{1\over b^3} \partial_\sigma b^3\partial_\sigma\phi +V'(\phi)=0
\eeq
If we demand that the solution at $\sigma=0$ be regular, we must have
$\dot\phi =0$ there. The real solution for $b$ will in general have at
least another point, $\sigma=\sigma_1$  where $b$ goes to zero.
 As HT analyse the solution, this point is in general singular, but
 they argue, not sufficiently singular to disallow its use as a WKB
solution to the equations. $\phi$ diverges to infinity (logarithmically
in $\sigma-\sigma_1$) , and $b$ goes to zero. As an example, I will
examine the (unphysical) case where the potential is a linear
potential. The equation for $\phi$ is then
\beq
\phi={1\over 6} V'\int {1\over b^3}(\int  {  b^3}dt) dt
\eeq
with $b$ satisfying
\beq
(\dot b)^2= {1} -{1\over 6}  b^2( -(\dot\phi)^2 +V_0+V'(\phi))
\eeq
Thus $\dot\phi$ diverges as ${1\over b^3}$.Let $b(\sigma)\propto (\sigma_1-\sigma)^\alpha$,
 which gives 
\beq
\dot \phi = {A \over (\sigma_1-\sigma)^{3\alpha}} +B +...)
\eeq
and 
\beq
\phi= -{A\over(3\alpha-1)}(\sigma_1-\sigma)^{3\alpha-1}+C
\eeq
Substituting into the equation for $b$ gives
\beq
D^2\alpha^2 (\sigma_1-\sigma)^{2(\alpha-1)} +1 
- D^2 (\sigma_1-\sigma)^{2(\alpha)}
 \left( {A^2 \over (\sigma_1-\sigma)^{6\alpha}} 
+V'({-A\over(3\alpha-1)(\sigma_1-\sigma)^{3\alpha-1}})\right)=0
\eeq
Assuming that $\alpha$ lies between 0 and 1, the second term in brackets
 proportional to $V'$ is smaller than the first term derived from
 $\dot\phi^2$ in the constraint equation. (This would be expected to be true of any potential which
 increased at a rate slower than exponential). Retaining only lowest
 order terms, we get $\alpha=1/3$. The derivative of the scalar field
and the curvature thus diverge as $1\over (\sigma_1-\sigma)$, and the
scalar field diverges logarithmically, in agreement with HT.

Of interest later will be the behaviour of this solution around
$\sigma=0$. Firstly, in order that the solution give a regular geometry
at $\sigma=0$, we must have $b(\sigma)=\sigma$. The
above equations are symmetric under the replacement of $b$ with $-b$.
The equations are also symmetric in $\sigma$ around $\sigma=0$. We can
therefor choose (and in fact must if the solution is to be analytic)
our solution for $b$
to be antisymmetric about $\sigma=0$, so that $b(-\sigma)=-b(\sigma)$. The Taylor
series expansion of $b(\sigma)$ will thus be a series in odd powers of
$\sigma$.  The solution for $\phi(\sigma)$ will be even in $\sigma$ since $\phi(0)$ is non-zero while $\dot\phi(0)$ is zero. Thus a Taylor series expansion of $\phi$ will be an expansion in even powers of $\sigma$.

Now, how do HT use this solution? They first match this Euclidean
solution to a Lorentzian solution across the $\psi=\pi/2 $ surface. At
this surface the extrinsic curvature of the slice is zero, at least everywhere except at $\sigma=\sigma1$, the singularity. They match
this Euclidean solution to the Lorentzian solution
\beq
ds^2= d\sigma^2+b(\sigma)^2(-d\tau^2 +cosh(\tau)^2 d\Omega^2)
\eeq
at $\tau=0$ which is also a slice with zero extrinsic curvature, and with 
the same 3-geometry,
\beq
ds_{(3)}^2= d\sigma^2 +b(\sigma)^2(d\Omega^2)
\eeq
 allowing a matching without the introduction of delta function
stresses at the matching surface.  The $\sigma=0$ surface in the above
$\sigma,\tau$ coordinates  is a null surface, similar to the horizon in
Rindler coordinates or the  Schwartzschild solution.  Define 
\ba
\Sigma= \int_0^\sigma {d\sigma'\over b(\sigma')}
\ea
and
\ba
U=e^{-\tau+\Sigma}\\
 V=-e^{\tau+\Sigma}.
\ea
Since near $\sigma=0$,  $b(\sigma)$ is odd with leading term equal to
$\sigma$, $\Sigma$ will be of the form of  a leading order divergence of $ln(\sigma)$
plus a series in even powers of $\sigma$.  In $U,~V$ coordinates we
have
\beq
ds^2={e^{2\Sigma}\over b(\sigma)^2} ( -dUdV +{1 \over 4} \left({U-V}\right)^2(d\Omega^2))
\eeq

From the expression for $b(\sigma)$ and $\Sigma$ near $\sigma=0$, the overall 
multiplier approaches unity as $\sigma$ goes to zero. We can solve for $\sigma$ as a function of $UV$ by noticing that 
\beq
-UV= e^{2\Sigma(\sigma)}= \sigma^2 f(\sigma^2)
\eeq
where $f(0)=1$ and $f$ has  a power series expansion in 
powers of $\sigma^2$. I will assume that we can invert this equation so that we can write $\sigma^2$ as a power series in $UV$ with leading term of just $UV$. Thus we will have $\sigma(UV)= \sqrt{-UV}(1+ ({\rm series ~in~ }UV)).$
The expression $\Sigma(\sigma)-\ln(b(\sigma))$ will be a power
series in $UV$ with lowest order term of unity. These arguments are formal, and may fail if the formal power series has zero radius of convergence. I will assume that this does not occur.

 Now define
\beq
F(UV)= e^{2\Sigma-2\ln(b(\sigma))}=-{UV \over b(\sigma(\zeta))^2}
\eeq
which, by the above arguments,   has a power series
 expansion in $UV$ and has
$F(0)=1$.
The metric is thus
\beq
ds^2= F(UV)(- dUdV +{1\over 4}(U-V)^2 d\Omega^2)
\eeq
By our assumption that the series solution for $F$ has a non-zero radius of convergence,  we can continue $F(UV)$ to positive values of $UV$ .   $F$  will
 be a real function for real values of $UV$.
 
Let me now  define new coordinates $t,\chi$ by
\ba
U=e^{T-\chi}\\
V=e^{\chi+T}
\ea
This gives
\beq
ds^2= F(e^{2T})e^{2T}( -dT^2 +d\chi^2 + \sinh^2(\chi)d\Omega^2)
\eeq
which is just the equation for an open spatially 
homogeneous universe with scale factor $e^T \sqrt{F(e^{2T})}$ .

Having examined the geometry, we can also examine the solution for the scalar field $\phi$. As argued above, $\phi$ has a Taylor series in even powers of $\sigma$ and thus has a Taylor series in powers of $UV$. It can thus also be continued in a regular fashion into the
$T,~\chi$ region, where it will be a function only of $T$, and independent of $\chi$.

Does this now mean that we have a an interpolation 4-geometry for a
homogeneous hyperbolic spatial three-geometry $T={\rm const}$? This
would be true if we could arrange for this hyperbolic spatial geometry
to be the only boundary of the four dimensional spacetime  given above.
Unfortunately, this is not the case. In particular, there is a
space-like boundary at infinity where $\tau$ goes to infinity for any
constant $\sigma$ (ie, in the region where $UV$ is negative). Ie, in order that this four geometry be regardable
as the interpolating geometry for some space-like 3-geometry, we
must find such a 3-geometry which can
 be chosen as the only boundary of this four geometry. To do so the
space-like surface will have to penetrate the horizon and enter the
region in which the $\tau,\sigma$ coordinates are defined, and
furthermore, it will have to include the singularity at
$\sigma=\sigma_1$.

 Figure 1 shows this solution in $U,~V$ coordinates. The dotted line at
U=V corresponds to $\chi=0$ the origin of the spherically symmetric
$\chi,\theta,\varphi$ coordinate system. The dashed line at
$UV=-e^{\Sigma_1}$ is the singularity where $\phi$ and $\dot\phi$ go to
infinity. I have drawn in a space-like hypersurface which for much of
it is a part of the hyperbolic spatially homogeneous geometry.
It then ceases to follow that geometry and enters the $\tau,\sigma$
region finally ending of the singularity. The 4-geometry presented
in HT {\bf is} a valid interpolating geometry for this spatial hypersurface (if we neglect the problem of the singularity).
Under the rules of the ``wave function of the universe" prescription,
this 4-geometry can be used to calculate the probability of this
three geometry occurring by looking at the imaginary part of the
action, as HT do.  

However, this 3-geometry, together with the
interpolating 4-geometry, is a singular geometry. At
$\sigma=\sigma_1$ ($-UV=\Sigma_1$), the
scalar field and the curvature diverge. Note that   this   
time-like singularity does not engulf the whole spacetime, as it does in
the Vilenkin model with a similar naked time-like singularity in an
asymptotically flat space-time\cite{Vilenkin}.

\epsfysize=3in
 \centerline{ \epsfbox[20 200 600 800]{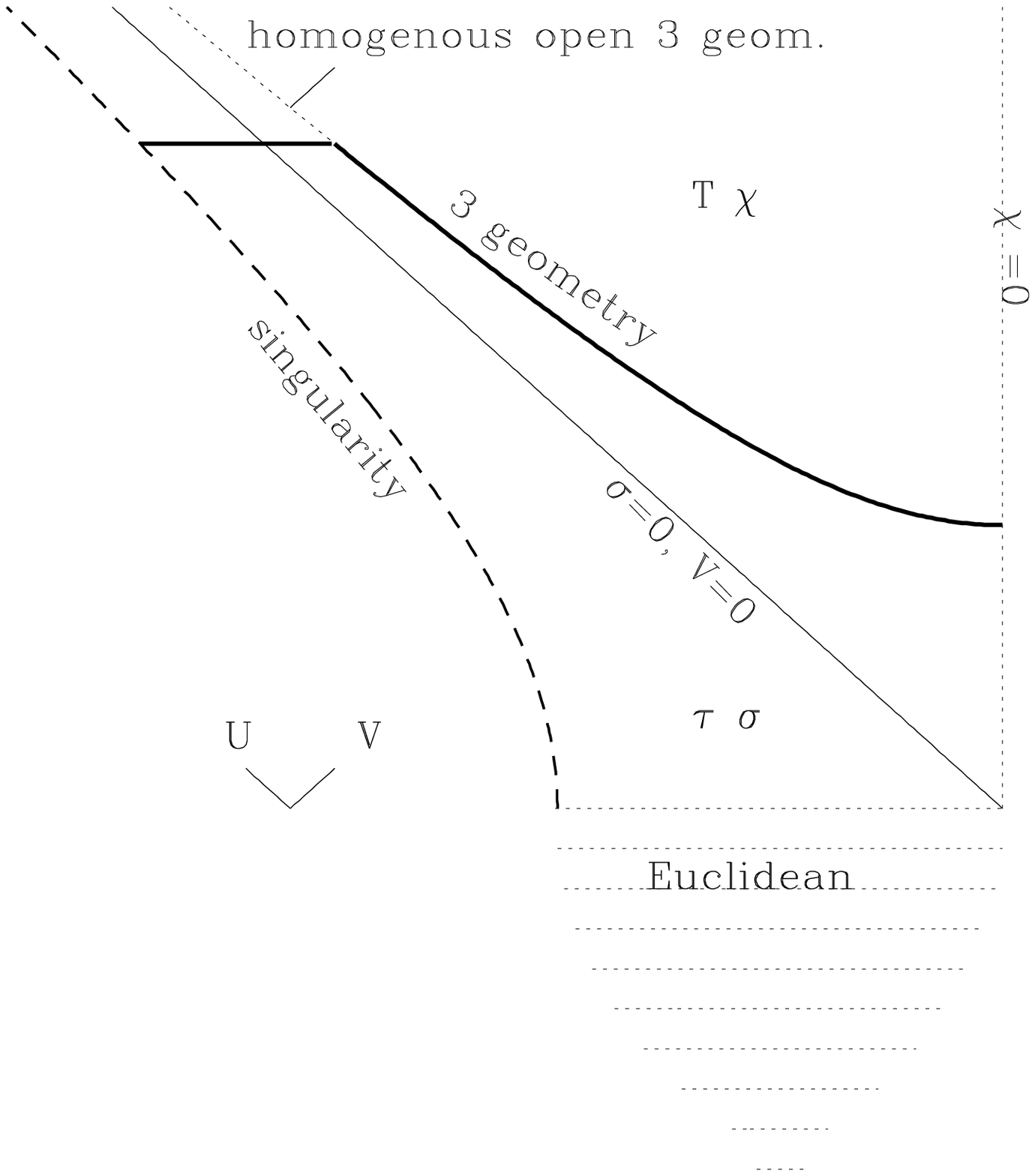}}
\centerline{Figure 1}
\centerline{ \it The global structure of the Hawking Turok Space-time}

This analysis thus answers a frequently asked question regarding the HT
solution, namely how can a finite compact Euclidean instanton create an
infinite open universe? The answer is that it doesn't. The instanton
creates a finite bounded universe, of which a part is a part of a
homogeneous hyperbolic geometry. Note that one can   make the open
universe section as large a part of the spatial section as one desires by
allowing the curve representing the three geometry to follow the open
universe for as long as one wishes.  Eventually however one must close
off the universe with the singularity.

 The singularity is a time-like attractive singularity (ie, geodesics
 fall toward and are not repelled away from the singularity) 
as we can see by looking
at the path $\sigma,~\theta,~\varphi$ all constant, which corresponds
to the particle remaining a constant proper distance from the
singularity (even though the circumferential distance is increasing
exponentially). The unit tangent vector is 
\beq
v^\mu= [ 0,1/b(\sigma), 0,0]. 
\eeq
Near $\sigma=\sigma_1$, the solution for $b$ is
 $b(\sigma)=A(\sigma_1-\sigma)^{1\over 3} $ 
where A is some constant. The acceleration vector for this path is 
\beq
{Dv^{a}\over Ds}= \left[ {A^2\over 3(\sigma_1-\sigma)},0,0,0 \right]
\eeq
which is an outward (away from the singularity) directed acceleration.
Ie, in order to keep the particle at a constant $\sigma$ (constant
proper distance from the singularity) a force directed away from the
singularity must be applied to the particle. A geodesic will thus tend
to decrease its proper distance from the singularity, which is why I
call it attractive. The singularity is also spherically symmetric, and
$\tau$-time dependent, although the scalar field is $\tau$-time
independent.
 
The question one must now raise is whether or not this solution is
actually what is wanted? Does one want a solution which singular? Does
one want a 3-geometry which is singular? Is the probability that
for producing an open universe bubble, or that for producing the
singularity, and how would one answer this question?

In short, HT calculate the probability for forming a closed  universe
which contains a part of a hyperbolic spatially homogeneous universe,
and also contains a naked, time-like, attractive singularity.  Whether
or not this thus corresponds to the creation of a closed universe is
open to question.

The chief problem with time-like singularities is that they have no
boundary conditions on the singularity which disallows matter from
streaming out of the singularity. Let $\Phi$ be some massless conformally invariant scalar
field (not related to $\phi$, the inflaton field).  
 The metric is  conformally
flat metric (as can be easily seen if we define $t=(U+V)/2$ and
$r=(V-U)/2$). The equation for $\Phi$ in the $U,~V$ coordinates
for a spherically symmetric solution is
\beq
\partial_U ( (U-V)^2 \partial_V (\sqrt{F}\Phi)) + 
\partial_V ((U-V)^2 \partial_U (\sqrt{F}\Phi))=0
 \eeq
 (The $\sqrt{F}$ term comes through the conformal transformation of
 $\Phi$ and the metric to the flat metric.) We can now choose   initial
 conditions on a $U+V=$const hypersurface such that $\sqrt{F}\Phi$ and
 $\partial_{U+V}\sqrt{F}\Phi$ are zero for all $U<U_i$ on this surface. Since the characteristics  of this equation are just the $U$ or $V$ constant lines,
 the complete solution will   be zero for $U<U_i$ everywhere, (at least until the $U=U_i$ line hits the $U=V$ origin of the spherical symmetry).
 Furthermore, we note that
 \beq
(UV)= e^{2\Sigma} = e^\Sigma_1 (1+ C(\sigma_1-\sigma)^{2/3}+...)
\eeq
or
\beq
(UV)-e^{2\Sigma_1} \propto (\sigma_1-\sigma)^{2/3}
\eeq
Thus 
\beq
F(UV)=UV/b(\sigma^2) \propto {e^{2\Sigma_1}\over (UV-e^{2\Sigma_1})}
\eeq
Thus, since the function $\sqrt{F}\Phi$ can be chosen to be regular
near the singularity, $\Phi$ will go to zero at the singularity as
$\sqrt{UV-e^{2\Sigma_1}}$.
 Ie,the solution will be such that the the wave $\Phi$ is emitted by
 the singularity, and that value of $\Phi$ is zero at the time-like
 singularity
(although its derivative does diverge). This of course does not take
into account the back reaction of this wave on the geometry of the
singularity, but it is difficult to see how that would rule out such
solutions.   Ie,
the time-like singularity
 can act as the source for matter streaming out into the universe.

In conclusion, the HT solution is not the calculation of the
wave-function at a homogeneous open universe, but rather of a universe
with an arbitrarily large section of such an open universe, but then
inhomogeneously closed off by a naked time-like singularity. Whether or
not such a solution is a valid one for the wave function of the
universe is an open question.

\end{document}